\documentclass[preprint,notoc]{JHEP3} 

\JHEPspecialurl{http://jhep.sissa.it/JOURNAL/JHEP3.tar.gz}

\usepackage{epsfig,multicol}

\newcommand\fverb{\setbox\pippobox=\hbox\bgroup\verb}
\newcommand\fverbdo{\egroup\medskip\noindent%
			\fbox{\unhbox\pippobox}\ }
\newcommand\fverbit{\egroup\item[\fbox{\unhbox\pippobox}]}
\newbox\pippobox

\def\to{\rightarrow}

\def\te{\tilde e}

\def\tb{\tilde b}

\def\tst{\tilde t}
\def\ttau{\tilde \tau}
\def\tmu{\tilde \mu}

\def\tw{\widetilde W}
\def\tz{\widetilde Z}
\def\alt{\stackrel{<}{\sim}}

\title{Neutralino relic density in minimal supergravity 
with co-annihilations}

\author{Howard Baer, Csaba Bal\'azs and Alexander Belyaev\\
	Department of Physics, Florida State University, 
                  Tallahassee, FL, USA 32306\\
	E-mail: \email{baer@hep.fsu.edu}, \email{balazs@hep.fsu.edu},
                 \email{belyaev@hep.fsu.edu}}
\preprint{January, 2002}

\preprint{\hepph{0202076}}	

\abstract{We evaluate the relic density of neutralinos in the minimal
supergravity (mSUGRA) model.  
All $2\rightarrow 2$ 
neutralino annihilation diagrams, as well as all 
processes involving sleptons, charginos%
, neutralinos and third generation squarks are included.
Relativistic thermal averaging of the velocity times cross sections is performed.
We find that co-annihilation effects are only important on the edges of the model
parameter space, where some amount of fine-tuning is necessary to obtain a 
reasonable relic density. Alternatively, at high $\tan\beta$, 
annihilation through very broad Higgs resonances gives rise to
an acceptable neutralino relic density over broad regions of parameter
space where little or no fine-tuning is needed. Finally, we compare
our results against the reach of various $e^+e^-$ and hadron 
colliders for supersymmetric matter.
}

\begin{document} 

\section{Introduction}

A wide variety of astrophysical measurements are being used to pin
down some of the basic cosmological parameters of the universe.
High resolution maps of the cosmic microwave background (CMB) 
radiation\cite{cmb}
imply that the energy density of the universe $\Omega=\rho/\rho_c \simeq 1$, 
consistent with inflationary cosmology. Here, $\rho_c= 3 H^2/8\pi G_N$
is the critical closure density of the universe, where $G_N$ is Newton's
constant and $H=100h$ km/sec/Mpc is the scaled Hubble constant.
The value of $h$ itself is determined to be $\sim 0.7\pm 0.1$ by
improved measurements of distant galaxies\cite{wlf}. 
Meanwhile, data from distant 
supernovae\cite{supernovae} imply a nonzero dark energy content of the universe
$\Omega_\Lambda \sim 0.7$, a result which is confirmed 
by fits to the CMB power spectrum\cite{jaffe}. Analyses of Big Bang
nucleosynthesis\cite{nucleo} imply the baryonic density 
$\Omega_b h^2\simeq 0.020\pm 0.002$, although the CMB fits suggest a 
somewhat higher value of $\sim 0.03$. Hot dark matter, for instance 
from massive neutrinos, should give only a small contribution to the
total matter density of the universe. In contrast, a variety of
data ranging from galactic rotation curves to 
large scale structure and the CMB imply a significant
density of cold dark matter (CDM)\cite{review} $\Omega_c h^2\simeq 0.2\pm 0.1$.

In many $R$-parity conserving supersymmetric
models of particle physics, the lightest neutralino ($\tz_1$) 
is also the lightest
SUSY particle (LSP); as such, it is massive, neutral and stable.
For this case,
relic neutralinos left over from the Big Bang provide an 
excellent candidate for the CDM content of the universe\cite{jkg}.
In the early universe, such neutralinos would exist in thermal equilibrium
with the cosmic soup. As the universe expanded and cooled, the thermal
energy would no longer be sufficient to produce neutralinos
at an appreciable rate, although they
could still annihilate away. Their number density is governed by the
Boltzmann equation formulated for a Friedmann-Robertson-Walker universe.

In this paper, our goal is to present results of calculations of
the neutralino relic density within the context of the paradigm
minimal supergravity model (mSUGRA, or CMSSM)\cite{sugra}. In mSUGRA, it
is assumed that SUSY breaking occurs in a hidden sector of the model, 
with SUSY breaking effects communicated from hidden to observable sectors via 
gravitational interactions. The model parameter space is given by
\begin{equation}
m_0,\ m_{1/2},\ A_0,\ \tan\beta\ {\rm and}\ sign(\mu ) .
\end{equation}
Here, $m_0$ is the universal scalar mass, 
$m_{1/2}$ is the universal gaugino mass and $A_0$ is the universal
trilinear mass all evaluated at $M_{GUT}$, while $\tan\beta$ is the ratio
of Higgs field vevs ($v_u/v_d$), and $\mu$ is a supersymmetric Higgs mass term.
The soft SUSY breaking parameters,
along with gauge and Yukawa couplings, evolve from $M_{GUT}$ to $M_{weak}$
according to their renormalization group (RG) equations. At $M_{weak}$,
the RG improved 1-loop effective potential is minimized,
and electroweak gauge symmetry is broken radiatively. In this report, we 
implement the mSUGRA solution encoded in ISAJET v7.59\cite{isajet}.

There is a long history of increasingly sophisticated solutions
for the relic density of neutralinos in supersymmetric 
models\cite{early,ows,barb,gkt,gs,gg,bottino,dn,
leszek,an,bb,eg,bk,ellis,darksusy,fmw,ellis_co,an2,pallis,
leszek2,manuel,drees,santoso}. 
The key ingredient to 
solving the Boltzmann equation is to evaluate the thermally averaged
neutralino annihilation cross section times velocity factor.
Traditionally, the solution is made by expanding the annihilation 
cross section as a 
power series in neutralino velocity, 
so that angular and energy integrals
can be evaluated analytically. The remaining integral over temperature
can then be performed numerically. The power series solution is excellent 
in many regions of model parameter space because the relic neutralino 
velocity is expected to be highly non-relativistic. 

However, it was
emphasized by Griest and Seckel that annihilations may occur through
$s$-channel resonances at high enough energies\cite{gs} that a relativistic
treatment of thermal averaging might be necessary. 
Drees and Nojiri found that at large values of the parameter $\tan\beta$, 
neutralino annihilation can be dominated by $s$-channel scattering through
very broad $A$ and $H$ Higgs resonances\cite{dn}. 
The proper formalism for relativistic
thermal averaging was developed by Gondolo and Gelmini (GG)\cite{gg}, 
and was implemented in the code of Baer and Brhlik\cite{bb,bk}. 
Working within the framework of the mSUGRA model, 
it was found\cite{bb,bk,ellis,leszek2,manuel} that at large $\tan\beta$, 
indeed large new regions of model
parameter space gave rise to reasonable values for the CDM relic density. 
At large $\tan\beta$, the $A$ and $H$ resonances
are broad enough (typically 10-50 GeV) that even if 
the quantity $2m_{\tz_1}$ is
several partial widths away from exact resonance, there can still be
a significant rate for neutralino annihilation. 
Thus, in the mSUGRA model at low $m_0$ and $\tan\beta$, 
neutralino annihilation is dominated by $t$-channel
slepton exchange, and reasonable values of the relic density occur
only for relatively low values of $m_0$ and $m_{1/2}$. At high $\tan\beta$,
a much larger parameter space is allowed, owing to off-resonance neutralino
annihilation through the broad Higgs resonances.

In addition, there exist regions of mSUGRA model parameter space where
co-an\-ni\-hi\-la\-tion processes are important, and even dominant. 
It was stressed
by Griest and Seckel\cite{gs} that in regions with a higgsino-like LSP, 
the $\tz_1$, $\tw_1$ and $\tz_2$ masses become nearly degenerate, so that
all three species can exist in thermal equilibrium, and annihilate against 
one another. The relativistic thermal averaging formalism of GG was 
extended to include co-annihilation processes by Edsj\"o and Gondolo\cite{eg}, 
and was implemented in the DarkSUSY code\cite{darksusy} 
for co-annihilation of charginos and heavier neutralinos.

The importance of neutralino-slepton co-annihilation was stressed
by Ellis {\it et al.} and others\cite{ellis_co,an2,pallis,leszek2,manuel}. 
In regions of mSUGRA parameter
space where $\tz_1$ and $\ttau_1$ (or other sleptons) were nearly 
degenerate (at low $m_0$),
co-annihilations could give rise to reasonable values of the relic density
even at very large values of $m_{1/2}$, at both low and high $\tan\beta$.
In addition, for large values of the parameter $A_0$ or for non-universal
scalar masses, top or bottom squark masses could become nearly degenerate 
with the $\tz_1$, so that squark co-annihilation processes can become
important as well\cite{drees,santoso}. 

In this paper, our goal is to calculate the relic density of 
neutralinos in the mSUGRA model including co-annihilation processes
in addition to {\it relativistic} thermal averaging of 
the annihilation cross section times velocity. 
Since there are very many Feynman diagrams to evaluate for
neutralino annihilations and co-annihilations, we use 
CompHEP~v.33.24\cite{comphep}, which provides for fast and efficient 
automatic evaluation
of tree level processes in the SM or MSSM.
For initial states including
$\tz_1$, $\tz_2$, $\tw_1$, $\te_1$, $\tmu_1$, $\ttau_1$,
$\tst_1$ and $\tb_1$, we count 1722 subprocesses, 
including 7618 Feynman diagrams. For those processes we have 
calculated the squared matrix element and have written it down 
in the form of CompHEP {\it FORTRAN} output. 

The weak scale parameters from supersymmetric models are
generated using ISAJET v7.59, 
and interfaced with the squared matrix elements from CompHEP.
Details of our computational algorithm are given in Sec. 2. In Sec. 3
we present a variety of results for the relic density in mSUGRA model
parameter space. Much of parameter space is ruled out at low $\tan\beta$
since the relic density is too high, and would yield too small an age
of the universe. At high $\tan\beta$, large regions of parameter
space are available with a reasonable relic density in the
range $0.1< \Omega_{\tz_1} h^2 < 0.3$. In Sec. 4, we compare our results with
some previous results on the reach of colliders, and draw some
implications. In Sec. 5, we conclude.

As we were completing this work, the group of Belanger, Boudjema, Pukhov and
Semenov reported on a 
calculation similar to ours in scope and method\cite{belanger}.
In addition, a paper by Nihei, Roszkowski and de Austri appeared,
containing analytic calculations of all $\tz_1\tz_1$ 
annihilation cross sections\cite{roszkowski}.

\section{Calculational Details}

The evolution of the number density of supersymmetric relics in the
universe is described by the Boltzmann equation
as formulated for a 
Friedmann-Robertson-Walker universe. For calculations including 
many particle species, such as the case where co-annihilations
are important, there is a Boltzmann equation for each particle species.
Following Griest and Seckel\cite{gs}, the equations can be combined to obtain
a single equation
\begin{equation}
\frac{dn}{dt} =-3Hn-\langle\sigma_{eff}v\rangle\left(n^2-n_{eq}^2\right)
\end{equation}
where
\begin{equation}
n=\sum_{i=1}^{N} n_i
\end{equation}
and the sum extends over the $N$ particle species contributing to
the relic density, with $n_i$ being the number density of the $i$th
species.
Furthermore, $n_{eq,i}$ is the number density of the $i$th species 
in thermal equilibrium, given by
\begin{equation}
n_{eq,i} =\frac{g_im_i^2T}{2\pi^2}K_2\left(\frac{m_i}{T}\right) ,
\end{equation}
where $K_j$ is a modified Bessel function of the second kind of order $j$.

The quantity $\langle\sigma_{eff}v\rangle$ is the thermally averaged
cross section times velocity. A succinct expression for this quantity
using relativistic thermal averaging was computed by Gondolo and Gelmini
for the case of a single particle species\cite{gg}, and was extended by
Edsj\"o and Gondolo for the case including co-annihilations\cite{eg}.
We adopt this latter form, given by
\begin{equation}
\langle\sigma_{eff}v\rangle (x) =\frac{\int_2^\infty K_1\left({a\over x}
\right)
\sum_{i,j=1}^{N}\lambda(a^2,b_i^2,b_j^2)g_ig_j\sigma_{ij}(a)da}
{4x\left( \sum_{i=1}^{N}K_2\left({b_i\over x}\right) b_i^2 g_i\right)^2} ,
\end{equation}
where $x=T/m_{\tz_1}$ is the temperature in units of mass of the relic 
neutralino, $\sigma_{ij}$ is the cross section for the annihilation
reaction $ij\to X$ ($X$ is any allowed final state consisting of 
2 SM and/or Higgs particles), 
$\lambda(a^2,b_i^2,b_j^2) = 
a^4+b_i^4+b_j^4-2(a^2 b_i^2+a^2 b_j^2+ b_i^2 b_j^2)$, 
$a=\sqrt{s}/m_{\tz_1}$ and $b_i=m_i/m_{\tz_1}$.
This expression is our master formula for the relativistically
thermal averaged annihilation cross section times velocity.

To solve the Boltzmann equation, we introduce a freeze-out
temperature $T_F$, so that the relic density of neutralinos is given 
by\footnote{The procedure we follow gives numerical results valid
to about 10\% versus a direct numerical solution of the 
Boltzmann equation\cite{gg}.}
\begin{equation}
\Omega_{\tz_1} h^2= \frac{\rho (T_0)}{8.1\times 10^{-47}\ {\rm GeV}^4}
\end{equation}
where
\begin{equation}
\rho (T_0)\simeq 1.66 {1\over M_{Pl}}
\left(\frac{T_{m_{\tz_1}}}{T_\gamma}\right)^3 T_\gamma^3\sqrt{g_*}
\frac{1}{\int_0^{x_F}\langle \sigma_{eff}v\rangle dx} .
\end{equation}
The freeze-out temperature $x_F=T_F/m_{\tz_1}$ 
is determined as usual by an iterative
solution of the freeze-out relation
\begin{equation}
x_F^{-1}=\log\left[ \frac{m_{\tz_1}}{2\pi^3}\frac{g_{eff}}{2}
\sqrt{\frac{45}{2g_* G_N}}\langle\sigma_{eff}v\rangle (x_F) \, x_F^{1/2} \right] .
\end{equation}
Here, $g_{eff}$ denotes the effective number of degrees of freedom
of the co-annihilating particles, as defined by Griest and Seckel\cite{gs}.
The quantity $g_*$ is the SM effective degrees of freedom parameter
with $\sqrt{g_*}\simeq 9$ over our region of interest.

The challenge then is to evaluate all possible channels for
neutralino annihilation to SM and/or Higgs particles, and also
all co-annihilation reactions. The 7618 Feynman diagrams are evaluated
using CompHEP, leading to about 50 MB of {\it FORTRAN} code.
 To achieve our final result with relativistic thermal
averaging, a three-dimensional integral must be performed over
{\it i.}) the final state subprocess scattering angle $\theta$, 
{\it ii.}) the subprocess energy parameter $a=\sqrt{s}/m_{\tz_1}$,
and {\it iii.}) the temperature $T$ from freeze-out $T_F$ to the present
day temperature of the universe, which can effectively be taken to be 0.
We perform the three-dimensional integral using the BASES 
algorithm\cite{bases},
which implements sequentially improved sampling in multi-dimensional
Monte Carlo integration, generally with good convergence
properties.
We note that the three-dimensional integration
appearing in the case of our relativistic calculations involving
several species in thermal equilibrium is about 2 orders of magnitude more 
CPU-time consuming than the series expansion approach, 
which requires just one numerical integration.

\section{Results}

Our first results in Fig. \ref{plane_10} show regions of 
$\Omega_{\tz_1} h^2$ in the 
$m_0\ vs.\ m_{1/2}$ plane in the minimal supergravity model
for $A_0=0$, $\tan\beta =10$ and for $\mu<0$ and $\mu >0$. The upper
plots show the contribution if only $\tz_1\tz_1$ annihilation reactions
occur, while the lower frames include as well all co-annihilation processes.
The red shaded regions are excluded by theoretical
constraints (lack of REWSB on the right, a charged LSP in the upper left).
The unshaded regions have $\Omega_{\tz_1}h^2 >1$, and should be excluded, 
as they would lead to a universe of age less
than 10 billion years, in conflict with the oldest stars found in globular
clusters. 
The light blue shaded regions have $\Omega_{\tz_1} h^2 <0.02$, 
which wouldn't be 
enough CDM even to explain galactic rotation curves. The green region
yields values of $0.1< \Omega_{\tz_1} h^2 <0.3$, {\it i.e.} in the most 
cosmologically favored region. The yellow ($0.02< \Omega_{\tz_1} h^2 <0.1$) and
dark blue ($0.3< \Omega_{\tz_1} h^2 <1$) correspond to regions with 
intermediate values of low and high relic density, respectively. 
Points with $m_{1/2}\alt 150$ GeV give rise to chargino masses below
bounds from LEP2; the LEP2 excluded regions due to chargino, slepton and Higgs
searches are not shown on these plots, 
but will be shown in Sec. 4.

\FIGURE[t]{\epsfig{file=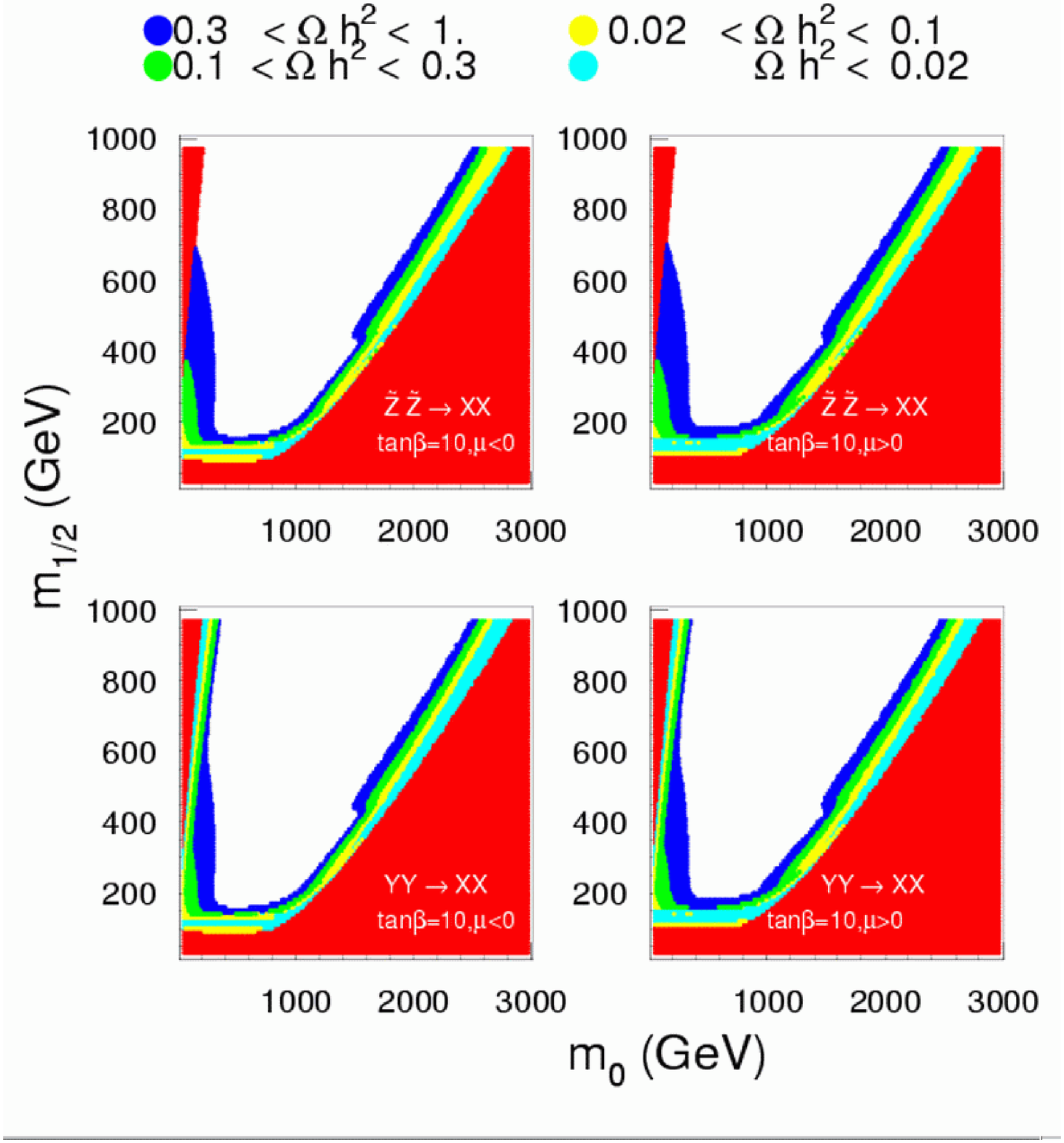,width=15cm} 
        \caption{Regions of neutralino relic density in the 
        $m_0\ vs.\ m_{1/2}$ plane for $A_0=0$ and $\tan\beta =10$.
        The upper two frames show the contribution for only $\tz_1\tz_1$ 
        annihilation, while the lower frames include as well all 
        co-annihilation processes.}
	\label{plane_10}}

The structure of these plots can be understood by examining
the thermally averaged cross section times velocity, integrated from
zero temperature to $T_F$. 
In Fig. \ref{1d_sp300_10} we show this quantity
for a variety of contributing subprocesses plotted versus $m_0$
for fixed $m_{1/2}=300$ GeV, $\mu >0$, and all other parameters as in 
Fig.~\ref{plane_10}. At low values of $m_0$, the neutralino annihilation
cross section is dominated by $t$-channel scattering into leptons pairs,
      as shown by the black solid curve. 
However, at the very lowest values of $m_0$,
the annihilation rate is sharply increased by neutralino-stau and stau-stau
co-annihilations, leading to very low relic densities where 
$m_{\tz_1}\simeq m_{\ttau_1}$\cite{ellis_co}. As $m_0$ increases, the
slepton masses also increase, which suppresses the annihilation
cross section, and the relic density rises to values $\Omega_{\tz_1}h^2>1$.
When $m_0$ increases further, to beyond the $\sim 1$ TeV level, 
and approaches the excluded region, the
magnitude of the $\mu$ parameter falls, and the higgsino component of
$\tz_1$ increases.
This is the so called ``focus point'' region, explored in
Ref. \cite{fmw}.
In this region, the annihilation rate
is dominated by scattering into $WW$, $ZZ$ and $Zh$ channels. 
At even higher $m_0$ values, $m_{\tz_1}\simeq m_{\tw_1}\simeq m_{\tz_2}$,
and these co-annihilation channels increase even more the annihilation rate.
Finally, at the large $m_0$ bound on parameter space, $|\mu |\to 0$,
and appropriate REWSB no longer occurs.
Most of the structure of Fig. \ref{plane_10} can be understood in these
terms, with the exception being the horizontal band of very low
      relic density at $m_{1/2}\simeq 125$ GeV. 

\FIGURE[t]{\epsfig{file=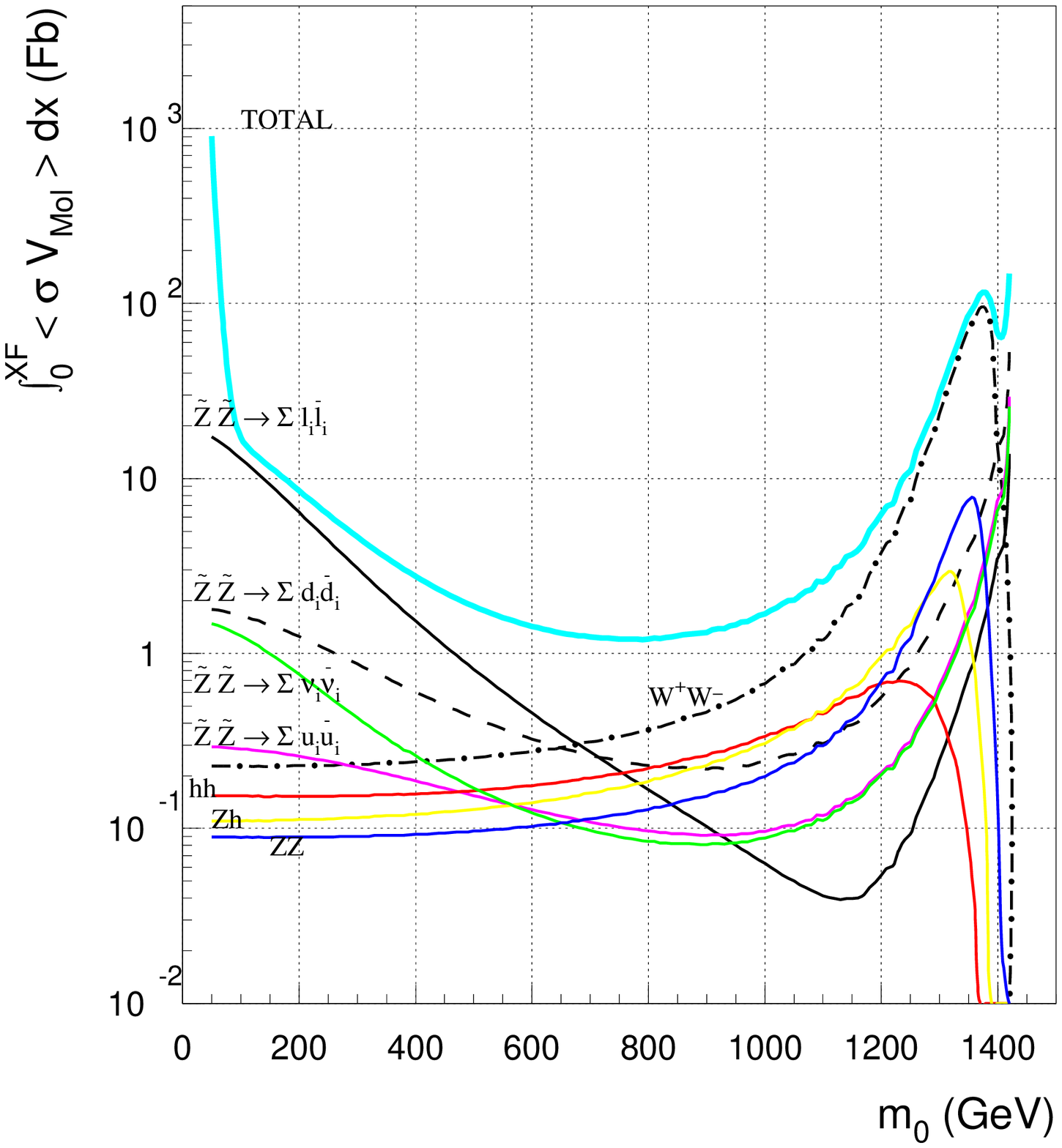,width=10cm} 
        \caption{Thermally averaged cross section times 
        velocity integrated from $T=0$ to $T_F$, for various component
        subprocess cross sections. The blue curve denotes the
        total of all annihilation and co-annihilation
        reactions. We show results 
        versus $m_0$ for $m_{1/2}=300$ GeV, $\mu >0$, $A_0=0$ 
        and $\tan\beta =10$.}%
	\label{1d_sp300_10}}
      
In this region, which is nearly
excluded by LEP2 bounds on the chargino mass, there is enhanced 
neutralino annihilation through the $Z$ and $h$ resonances. In fact, 
a higher degree of resolution on our plots would resolve 
these horizontal bands into {\it two} bands, corresponding to each of the 
separate resonances, as shown in Ref. \cite{bb}.
Finally, the glitch in contours around $m_0\sim 1500$ GeV 
and $m_{1/2}\sim 425$ GeV occurs because $m_{\tz_1}\simeq m_t=175$ GeV,
so that $\sigma (\tz_1\tz_1\to t\bar{t})$ becomes large.

The $m_0\ vs.\ m_{1/2}$ planes for $\tan\beta =30$ are shown in Fig. 
\ref{plane_30}. The structure of these plots are qualitatively
the same as in Fig. \ref{plane_10}. Quantitatively, they differ
mainly in that the cosmologically favored regions are expanding
as $\tan\beta$ grows. One reason is that the light stau becomes even
lighter as $\tan\beta$ increases, and this increases the neutralino
annihilation rate $\tz_1\tz_1\to \tau\bar{\tau}$ 
through $t$-channel stau exchange. In addition, the bottom and
tau Yukawa couplings increase with $\tan\beta$, which increases
the annihilation cross sections into $\tau$s and $b$s. 
Finally, the $H$ and $A$ Higgs boson masses are decreasing with $\tan\beta$,
and annihilation rates which proceed through these resonances increase. 
Co-annihilations again gives enhanced annihilation cross sections
on the left and farthest right hand sides of the allowed parameter space.

\FIGURE[t]{\epsfig{file=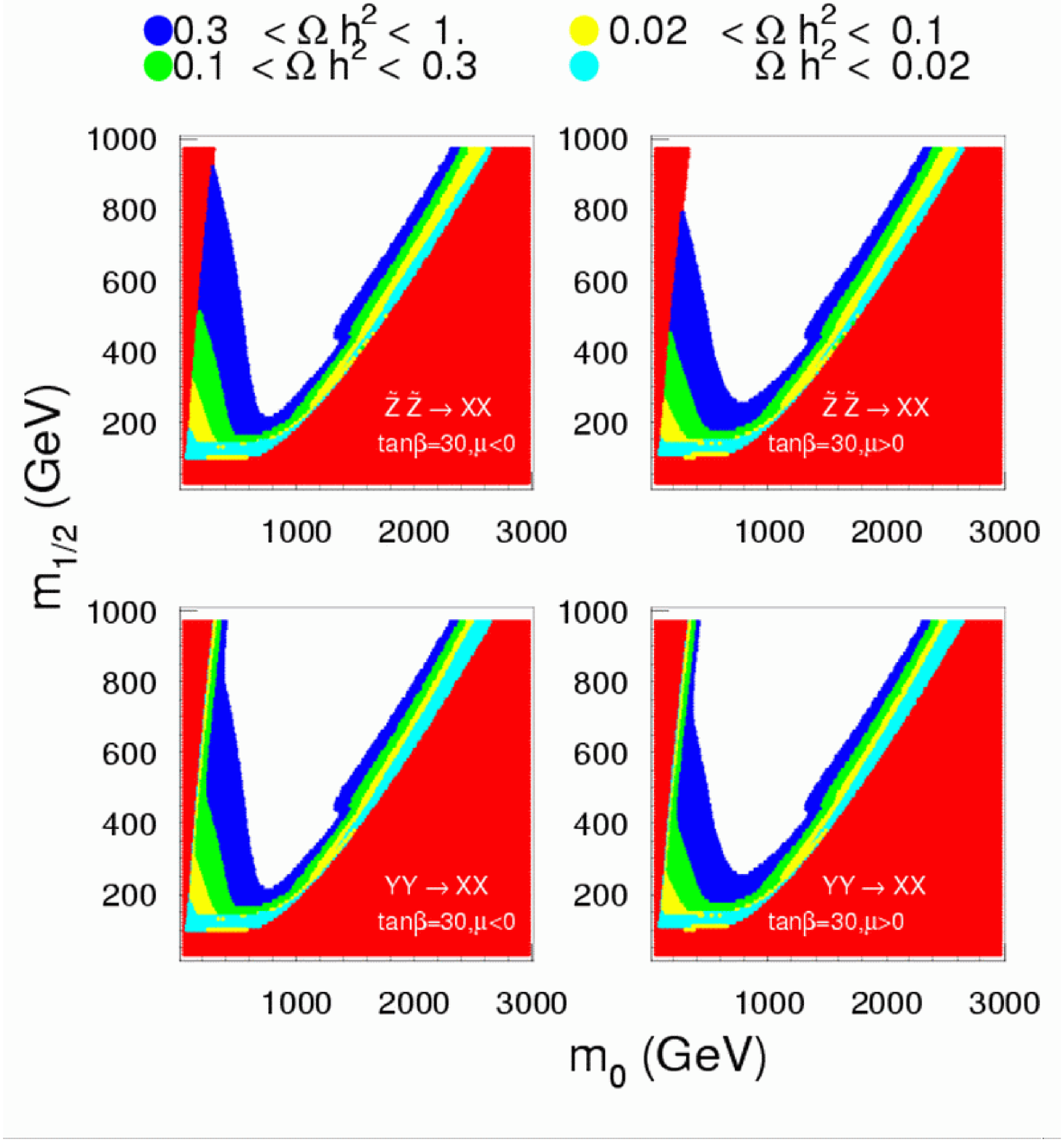,width=15cm} 
        \caption{Regions of neutralino relic density in the 
        $m_0\ vs.\ m_{1/2}$ plane for $A_0=0$ and $\tan\beta =30$.
        The upper two frames show the contribution for only $\tz_1\tz_1$ 
        annihilation, while the lower frames include as well all 
        co-annihilation processes.}
	\label{plane_30}}

In Fig. \ref{plane_45}, we show the $m_0\ vs.\ m_{1/2}$ plane for
$\tan\beta =45$. In this case, the structure of the plane is changing
qualitatively, especially for $\mu <0$. First, there is a new
region of disallowed parameter space for $\mu <0$ in the lower left
due to $m_A^2<0$, which signals a breakdown of the REWSB mechanism.
Second, a corridor of very low relic density passes 
diagonally through the plot. The center of this region is where
$2m_{\tz_1}\simeq m_A$ and $m_H$. At the $A$ and $H$ resonance, there
is very efficient neutralino annihilation into $b\bar{b}$ final
states. This is illustrated in Fig. \ref{1d_sp600}, where we show the
integrated annihilation cross section times velocity versus $m_0$
for $m_{1/2}= 600$ GeV and $\mu <0$. At the very lowest values of $m_0$, there
is again the sharp peak due to neutralino-stau and stau-stau
co-annihilations. For larger values of $m_0$, however, 
the annihilation rate is dominantly into $b\bar{b}$ final states
over almost the entire $m_0$ range. This is due to the large
annihilation rates through the $s$-channel $A$ and $H$ diagrams,
even when the reactions occur off resonance. In this case, the widths
of the $A$ and $H$ are so large (both $\sim 10-40$ GeV across the range
in $m_0$ shown) that efficient $s$-channel annihilation can occur
throughout the bulk of parameter space, even when the resonance condition
is not exactly fulfilled. The resonance annihilation is explicitly 
displayed in this plot as the annihilation bump at $m_0$ just
below 1300 GeV. Another annihilation possibility is that
$\tz_1\tz_1\to b\bar{b}$ via $t$ and $u$ channel graphs.
In fact, these annihilation graphs are enhanced due to the large $b$
Yukawa coupling and decreasing value of $m_{\tb_1}$, but we have checked
that the $s$-channel annihilation is still far the dominant channel.
Annihilation into $\tau\bar{\tau}$ is the next most likely channel, but
is always below the level of annihilation into $b\bar{b}$ for the 
parameters shown in Fig. \ref{1d_sp600}. At even higher values of 
$m_0$ where the higgsino component of $\tz_1$ becomes non-negligible,
the annihilations into
$WW$ and $ZZ$ again dominate; finally, at the highest values of
$m_0$, the $\tw_1$ and $\tz_2$ co-annihilation channels become important.

\FIGURE[t]{\epsfig{file=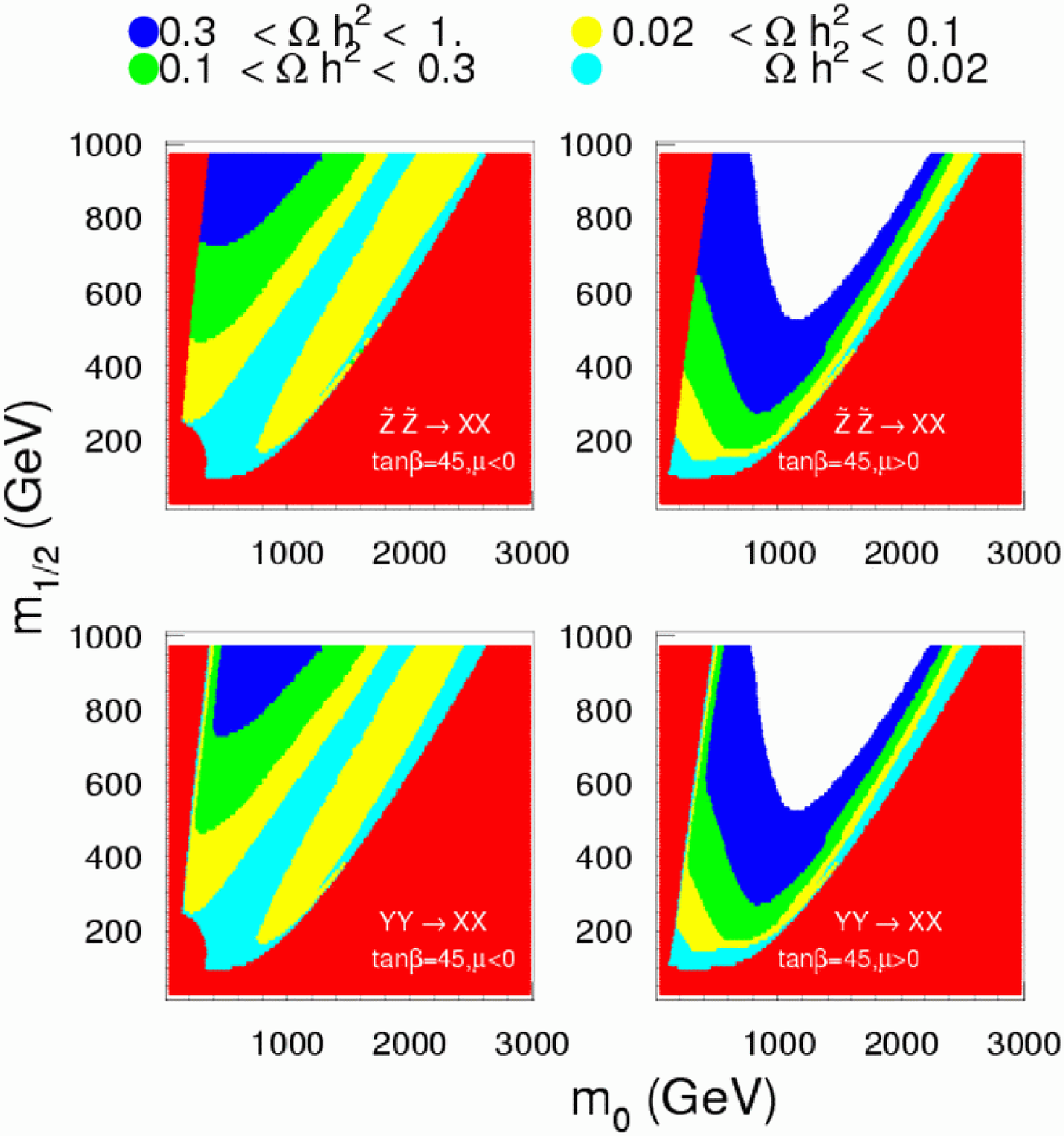,width=15cm} 
        \caption{Regions of neutralino relic density in the 
        $m_0\ vs.\ m_{1/2}$ plane for $A_0=0$ and $\tan\beta =45$.
        The upper two frames show the contribution for only $\tz_1\tz_1$ 
        annihilation, while the lower frames include as well all 
        co-annihilation processes.}
	\label{plane_45}}

\FIGURE[t]{\epsfig{file=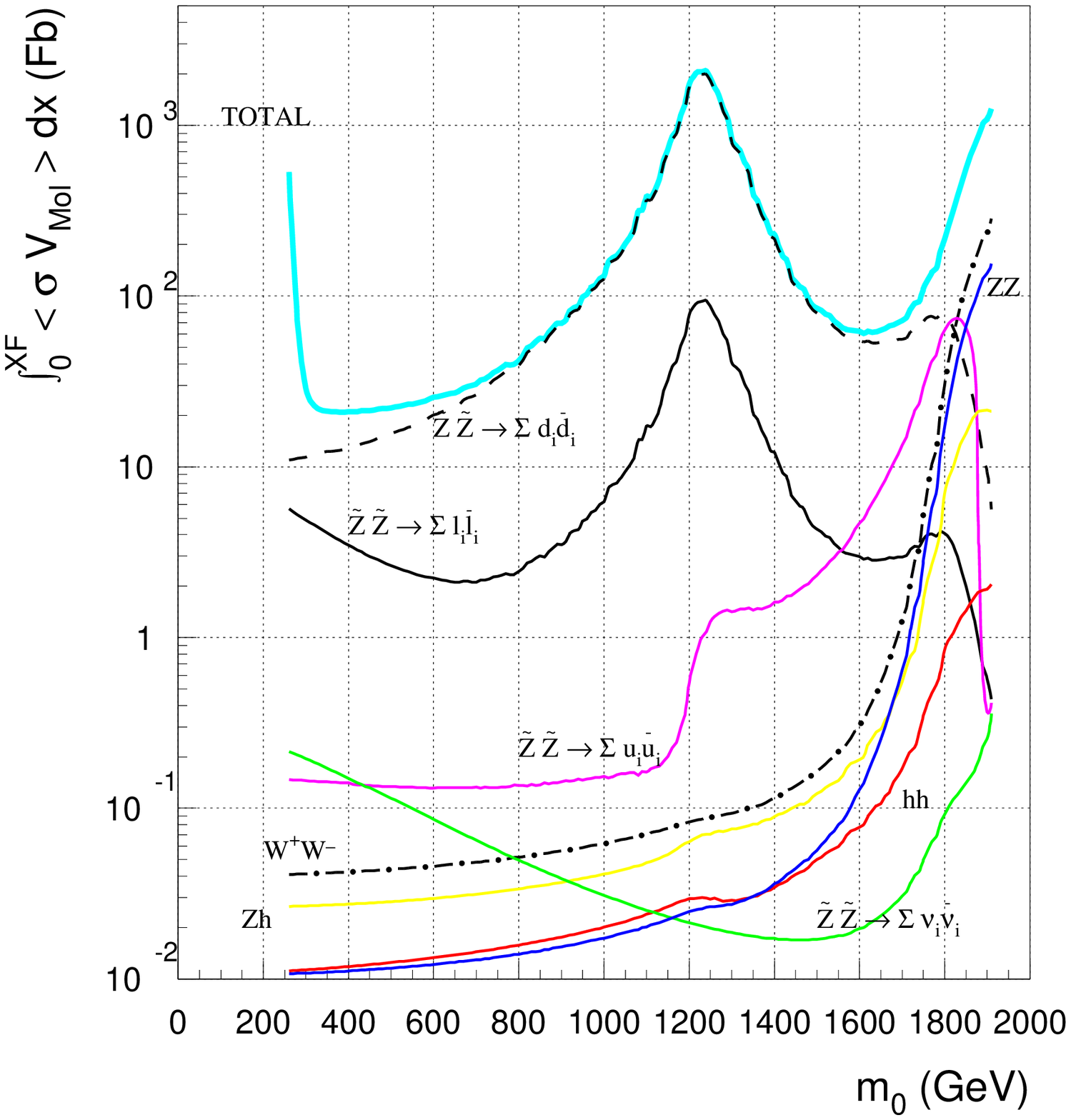,width=10cm} 
        \caption{Thermally averaged cross section times 
        velocity evaluated at $T_F$ for various component
        subprocess cross sections. The blue curve denotes the
        total of all annihilation and co-annihilation
        reactions. We show results 
        versus $m_0$ for $m_{1/2}=600$ GeV, $\mu <0$, $A_0=0$ 
        and $\tan\beta =45$.}%
	\label{1d_sp600}}

In Fig. \ref{1d_sp300}, we show again the subprocess annihilation rates
versus $m_0$ for $\tan\beta =45$, 
but this time for $\mu >0$ and for $m_{1/2}=300$ GeV. Although
no explicit resonance is evident for $\mu >0$, the dominant
annihilations are once again into $b\bar{b}$ final states over most of
the parameter space, due to the very wide Higgs resonances.
At the highest values of $m_0$, where $\mu$ is becoming small, 
the annihilation rate into the dominant $WW$ and $ZZ$ final states
becomes suppressed. The suppression is due to the diminishing mass
of the $\tz_1$ as $\mu \to 0$. As $m_{\tz_1}$ falls below 
$M_Z$ and then $M_W$, there is thermal suppression of annihilation
into the $ZZ$ and $WW$ final states.

\FIGURE[t]{\epsfig{file=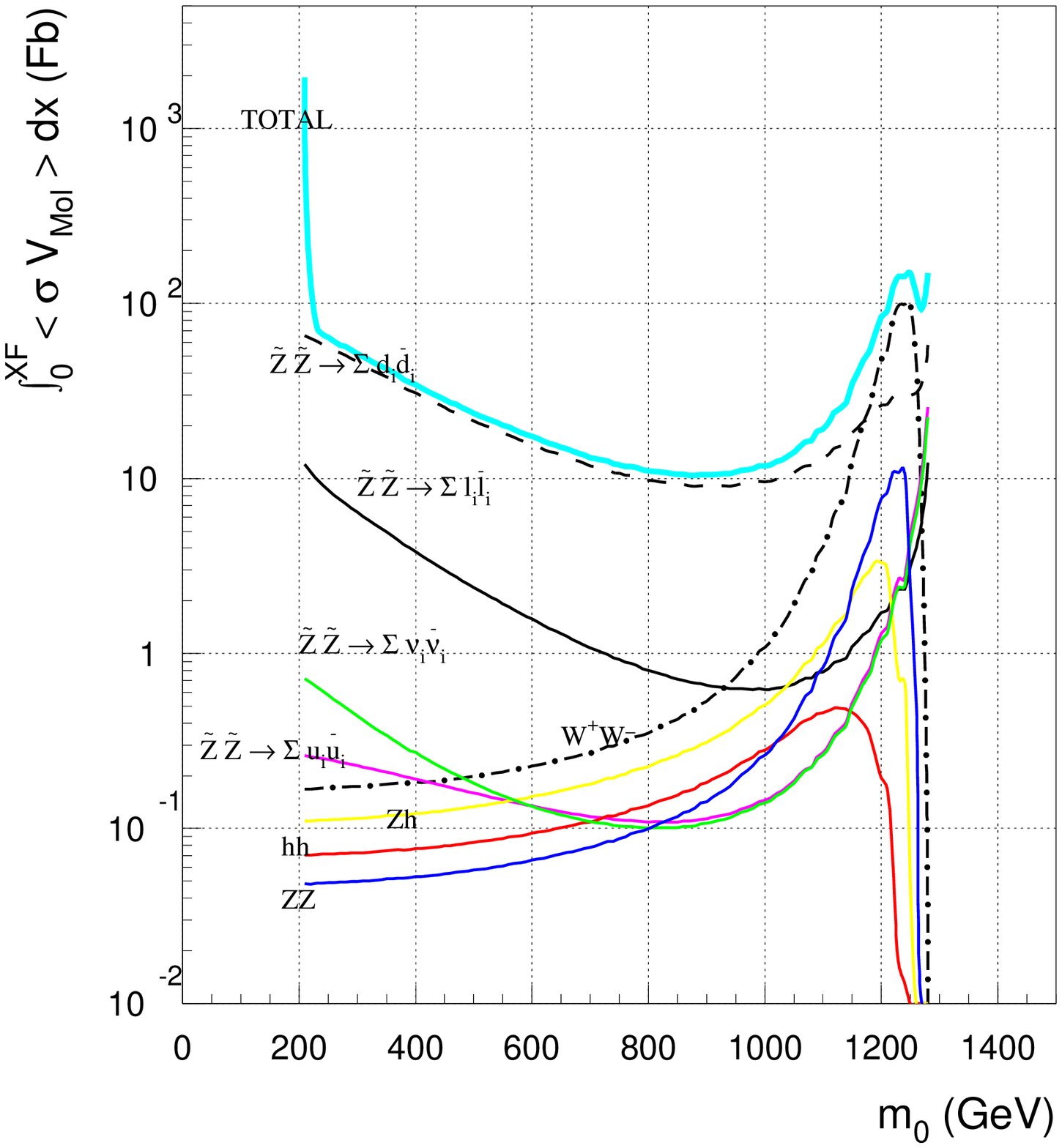,width=10cm} 
        \caption{Thermally averaged cross section times 
        velocity evaluated at $T_F$ for various component
        subprocess cross sections. The blue curve denotes the
        total of all annihilation and co-annihilation
        reactions. We show results 
        versus $m_0$ for $m_{1/2}=300$ GeV, $\mu >0$, $A_0=0$ 
        and $\tan\beta =45$.}%
	\label{1d_sp300}}

To summarize the regions of mSUGRA model parameter space with 
reasonable values of neutralino relic density, we can
label four important regions: $i.$) annihilation through $t$-channel
slepton-- especially stau-- exchange, as occurs for low values
of $m_0$ and $m_{1/2}$, $ii.$) the stau co-annihilation region for
low values of $m_0$ on the edge of the excluded region, $iii.$) the large
$m_0$ region with non-negligible higgsino-component
annihilation, and also $\tw_1$
(and possibly $\tz_2$) co-annihilation occurs near the
edge of the limit of parameter space, and $iv.$) annihilation into
$b\bar{b}$ and $\tau\bar{\tau}$ final states through $s$-channel
$A$ and $H$ resonances at high $\tan\beta$. Other regions can include
top or bottom squark co-annihilation for large values of $A_0$, again
on the edge of parameter space where $\tst_1$ or $\tb_1$ become light,
or annihilation through $Z$ or $h$ resonances. These latter regions
are essentially excluded now by constraints on sparticle masses from
LEP2.

It is useful to view the relic density $\Omega_{\tz_1}h^2$ directly as
a function of model parameters. We show in Fig. \ref{1d_600} the
value of $\Omega_{\tz_1}h^2$ versus the parameter $m_0$ for fixed
$m_{1/2}=600$ GeV, $A_0=0$, $\mu <0$ and for $\tan\beta =10$, 30 and 45.
The dashed curves show the result with no co-annihilations, while
the solid curves yield the complete calculation. The shaded band denotes the
cosmologically favored region with $0.1<\Omega_{\tz_1}h^2<0.3$. 
For this value of $m_{1/2}$, the lower $\tan\beta$ curves yield
a favored relic density only in the very low and very high $m_0$ regions,
and here the curves have a very sharp slope. The large slope is 
indicative of large fine-tuning, in that a small change of model parameters,
in this case $m_0$, yields a large change in $\Omega_{\tz_1} h^2$.
In contrast, the $\tan\beta =45$ curve shows a large region 
with good relic density and nearly zero slope, hence very little fine-tuning. 

\FIGURE[t]{\epsfig{file=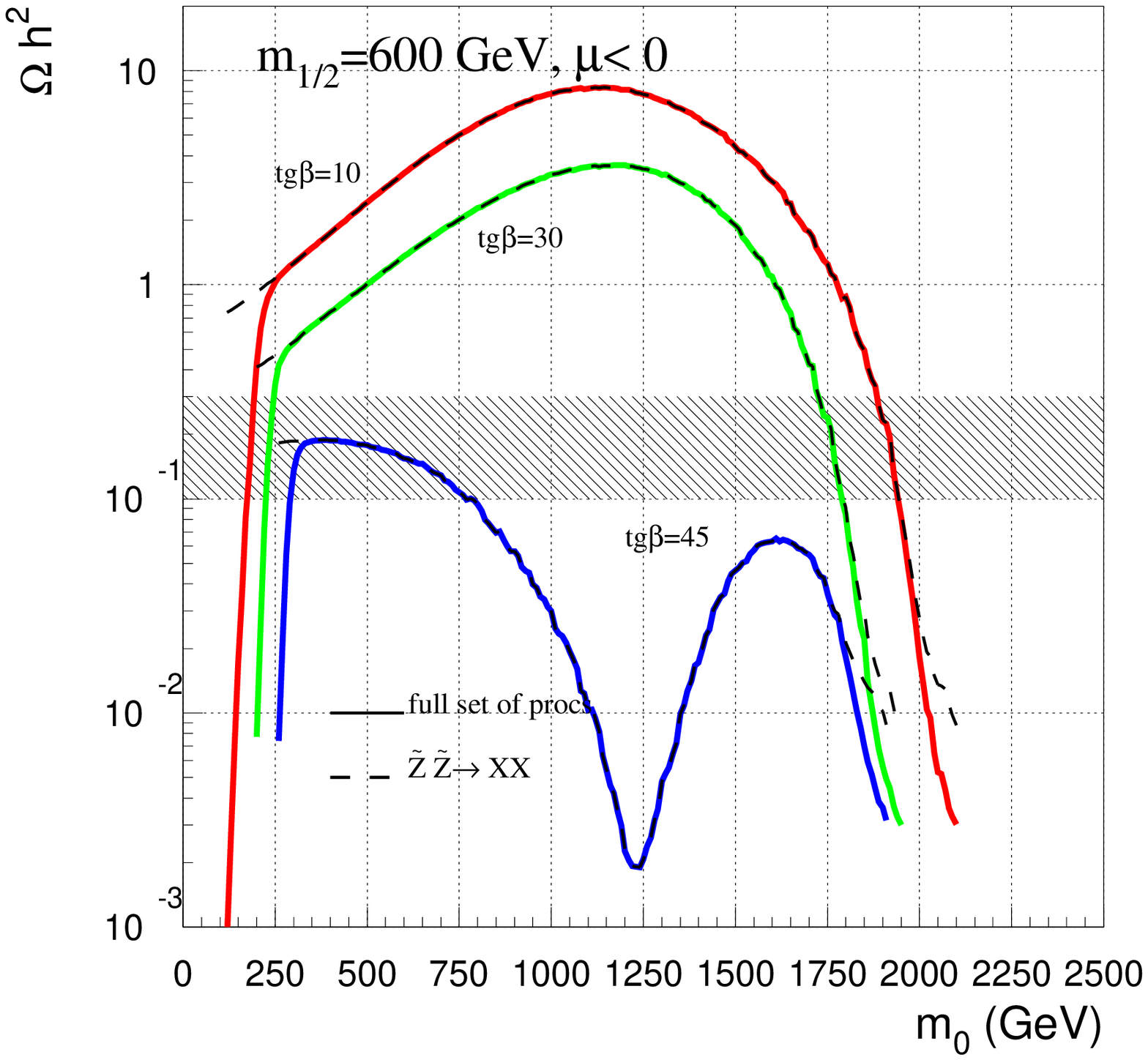,width=10cm} 
        \caption{Neutralino relic density $\Omega_{\tz_1} h^2$
        versus $m_0$ for $m_{1/2}=600$, $A_0=0$, $\mu <0$ and 
        $\tan\beta =10$, 30 and 45. The shaded region denotes
        the cosmologically favored values of $\Omega_{\tz_1} h^2$. }
	\label{1d_600}}

In Fig. \ref{ft_1}, we show the corresponding values of the fine-tuning,
basically the logarithmic derivative, as advocated by Ellis and 
Olive\cite{finetune}:
\begin{equation}
\Delta (m_0 )=\left|\frac{m_0}{\Omega_{\tz_1} h^2} 
\frac{\partial \Omega_{\tz_1} h^2}{\partial m_0}\right|
\label{Eq:FineTuning}
\end{equation}
%
As indicated earlier, the low fine-tuning regions mostly coincide with that of 
neutralino annihilation via $t$-channel slepton exchange (region $i.$)),
or off-resonance annihilation through $A$ and $H$ (region $iv.$)). 
The co-annihilation region $ii.)$ and focus point region $iii.$) tend to 
have higher fine-tunings due 
to the steep rise of the cross sections. Regions with simultaneous low fine-
tuning and preferred $\Omega_{\tz_1} h^2$ values are the 
best candidates for viable mSUGRA parameters.

\FIGURE[t]{\epsfig{file=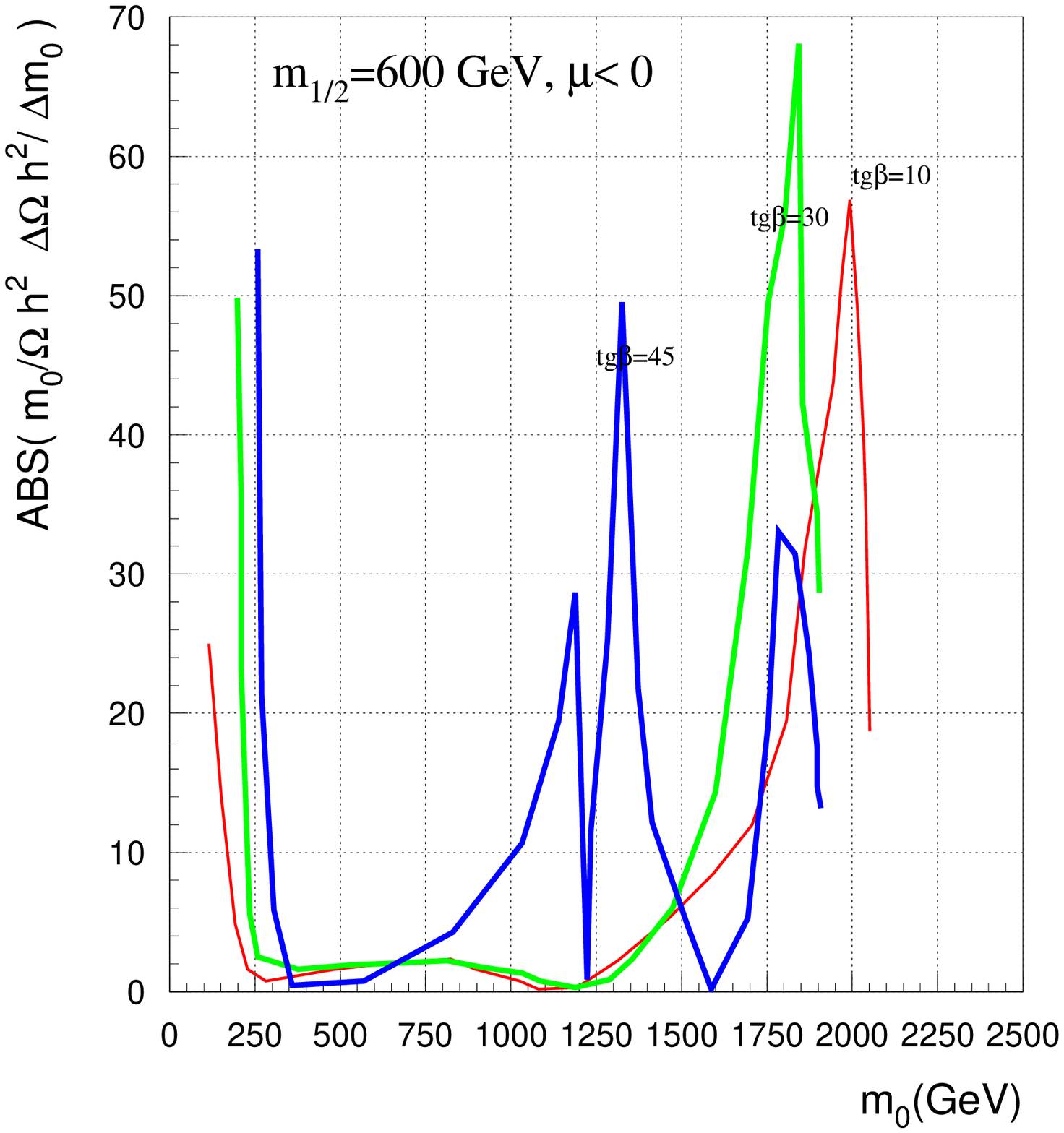,width=10cm}
        \caption{The fine tuning parameter 
        as the function of $m_0$ for $\tan\beta = 10,30,45$ and for the parameter
        slice $m_{1/2}=600$ GeV, $\mu < 0$.}
	\label{ft_1}}

In Fig. \ref{1d_300}, we show $\Omega_{\tz_1}h^2$ versus $m_0$ for
$m_{1/2}=300$ GeV, $A_0=0$, $\mu >0$ and the same three $\tan\beta$
parameters. 
The curves reflect the broad regions of parameter space with reasonable
relic density values at high $\tan\beta$.
The corresponding plot of the fine-tuning parameter
is shown in Fig. \ref{ft_2}. Again, there is large fine-tuning at the
edges of parameter space, but low fine-tuning in the intermediate
regions.
In conclusion, the relic density and the fine-tuning parameter together 
tend to prefer mSUGRA model parameters in regions $i.$) or $iv.$).
These two regions lead to distinct collider signatures for future
searches for supersymmetric matter.

\FIGURE[t]{\epsfig{file=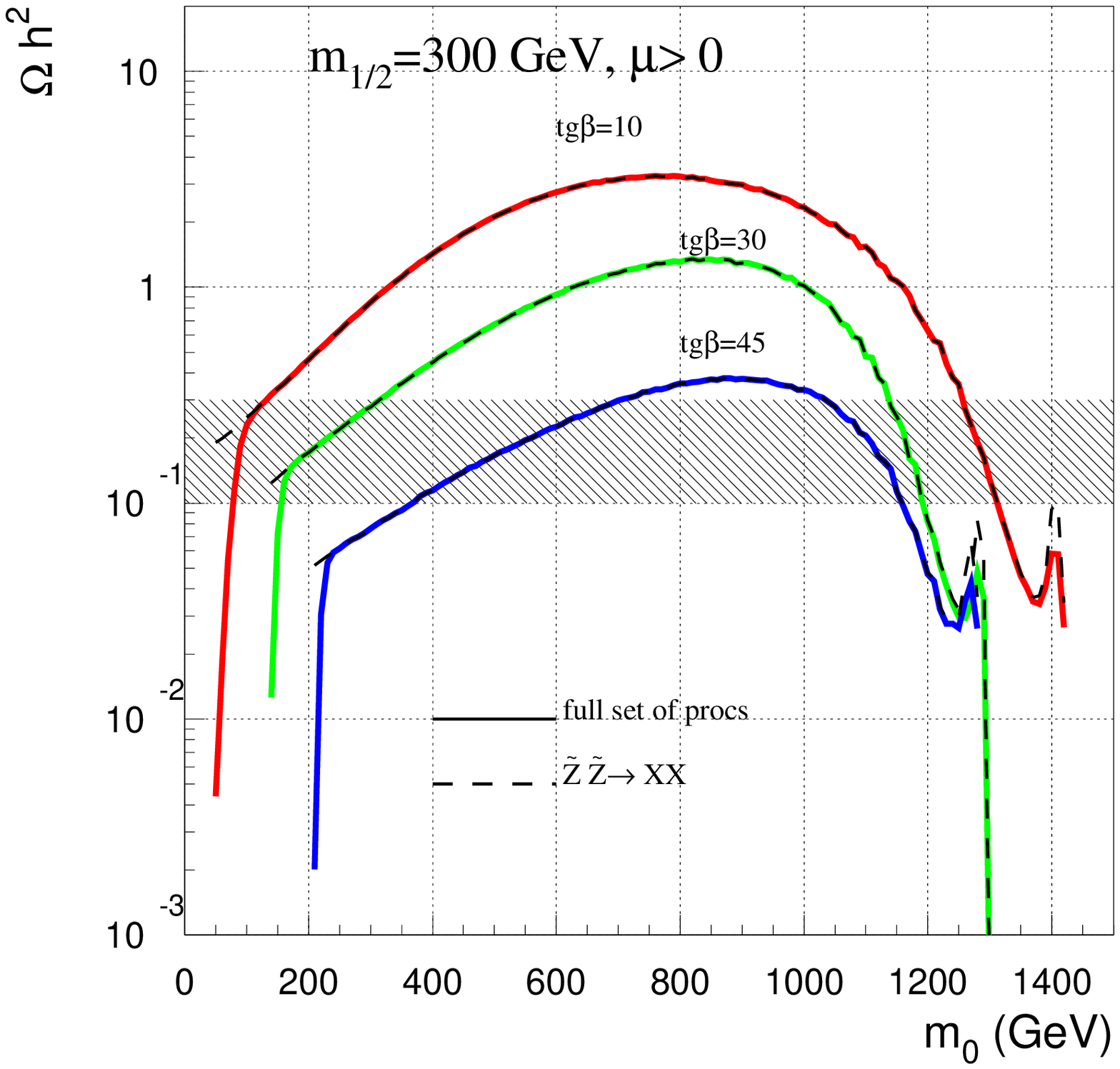,width=10cm} 
        \caption{Neutralino relic density $\Omega_{\tz_1} h^2$
        versus $m_0$ for $A_0=0$, $m_{1/2}=300$ GeV, $\mu >0$ 
        and $\tan\beta =10$, 30 and 45.}%
	\label{1d_300}}

\FIGURE[t]{\epsfig{file=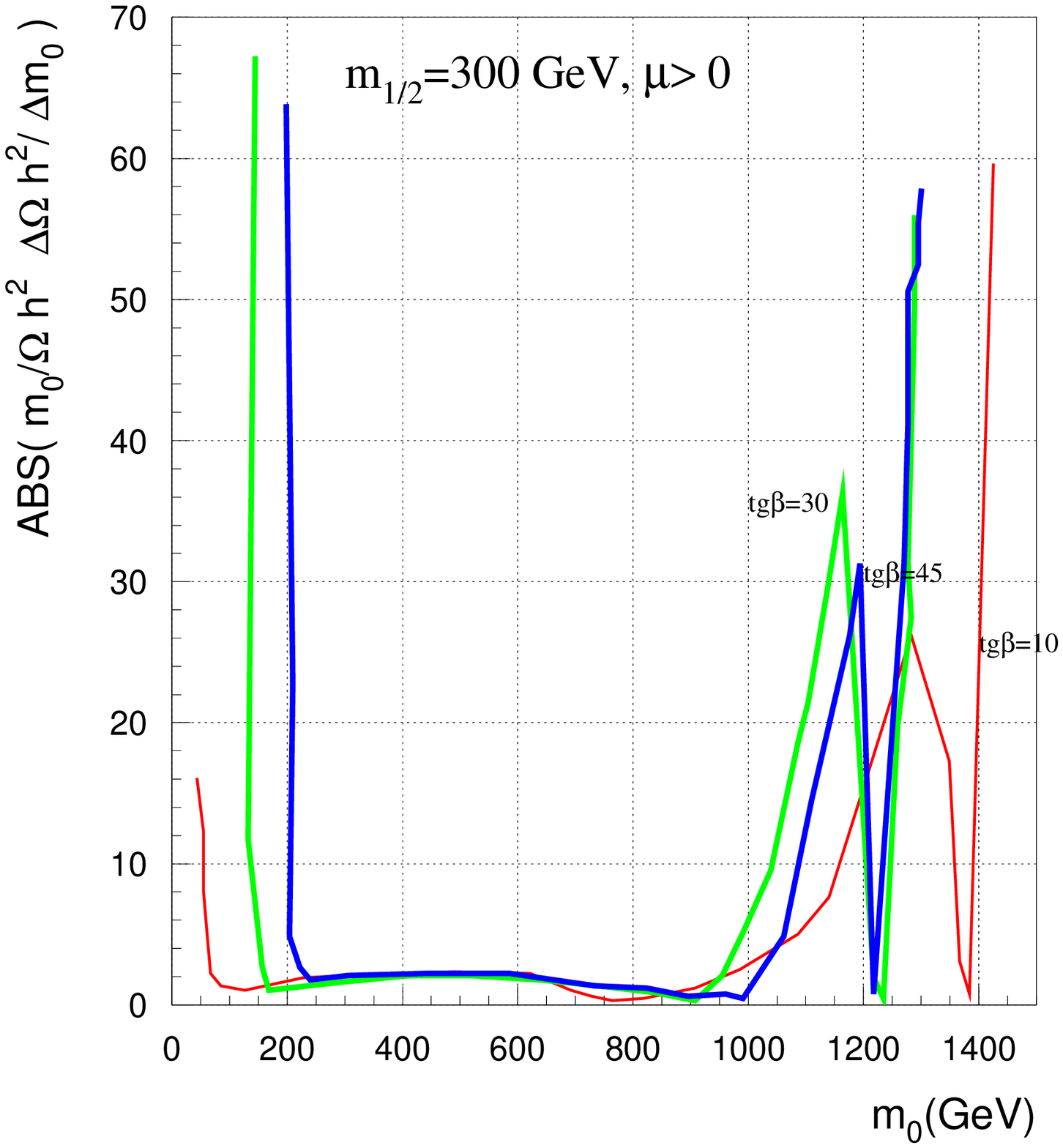,width=10cm}
        \caption{The fine tuning parameter 
        as the function of $m_0$ for $\tan\beta = 10,30,45$ and for the parameter
        slice $m_{1/2}=300$ GeV, $\mu >0$.}
	\label{ft_2}}

\section{Comparison with collider reaches}

It is worthwhile to compare our results on the neutralino
relic density with various collider 
reaches\footnote{Non-accelerator direct and indirect dark matter search
reach results are summarized in the last two papers of Ref. \cite{fmw}}. 
To do so, we first 
show in Fig. \ref{concl_10} the $m_0\ vs.\ m_{1/2}$ plane for
$\tan\beta =10$, $A_0=0$ and $\mu >0$, but this time
including information from different collider projections. 
First, the region excluded by LEP2 sparticle searches is shown by the pink
shading, and reflects mSUGRA model points where $m_{\tw_1}<100$ GeV, 
$m_{\te_1}<100$ GeV, or $m_{\ttau_1}< 76$ GeV\cite{lep2}. 
These LEP2 bounds sharply constrain the regions where neutralino annihilation
could occur via the $Z$ and light Higgs $h$ resonances. In addition, 
we plot contours of light Higgs boson mass $m_h=110$, 115 and 120 GeV.
Since the light Higgs scalar $h$ is usually SM-like in the mSUGRA model, 
the region below $m_h=115$ GeV is largely excluded\cite{lep2higgs}
by the direct LEP2 Higgs search. 
We note that the
Higgs mass varies slowly in parameter space, so a small change in Higgs mass
can lead to large changes in model parameters. Thus, these bounds
may have some fuzziness to them, reflecting uncertainties on the theoretical
calculations and experimental search results.
The reach of the Fermilab Tevatron for SUSY particles 
has been estimated recently in Ref. \cite{tevatron} for an integrated
luminosity of $25$ fb$^{-1}$. Almost all the reach comes from the 
search for clean trilepton events. The $3\sigma$ reach is denoted by the 
two lower black contours. The results show that Tevatron 
experiments will be able to probe a significant part of the 
favored relic density region where annihilation occurs through $t$-channel
slepton exchange. Also, some of the ``focus point'' region\cite{feng}
with large $m_0$ and small $m_{1/2}$ is accessible.

The reach of the CERN LHC is also shown for 10 fb$^{-1}$ of
integrated luminosity\cite{lhc}. The LHC reach extends well beyond the
$t$-channel slepton region, but cannot exclude all the low and high
$m_0$ regions corresponding to slepton co-annihilations or to
higgsino-like neutralinos, where annihilation cross sections are enhanced.
We remark, however, that these regions on the edge of parameter space, 
although perhaps not directly accessible to LHC searches, are also 
disfavored by fine-tuning requirements.

We also show the reach of a linear $e^+e^-$ collider for SUSY
particles for $\sqrt{s}=500$ GeV (NLC500) and $\sqrt{s}=1000$ GeV 
(NLC1000),
assuming 30 fb$^{-1}$ of integrated luminosity\cite{bmt}. The left-most
NLC region is explorable via slepton pair searches, while the lower and right
NLC regions are explorable via chargino pair searches. A small intermediate
region is accessible via $e^+e^-\to\tz_1\tz_2$ searches.

\FIGURE[t]{\epsfig{file=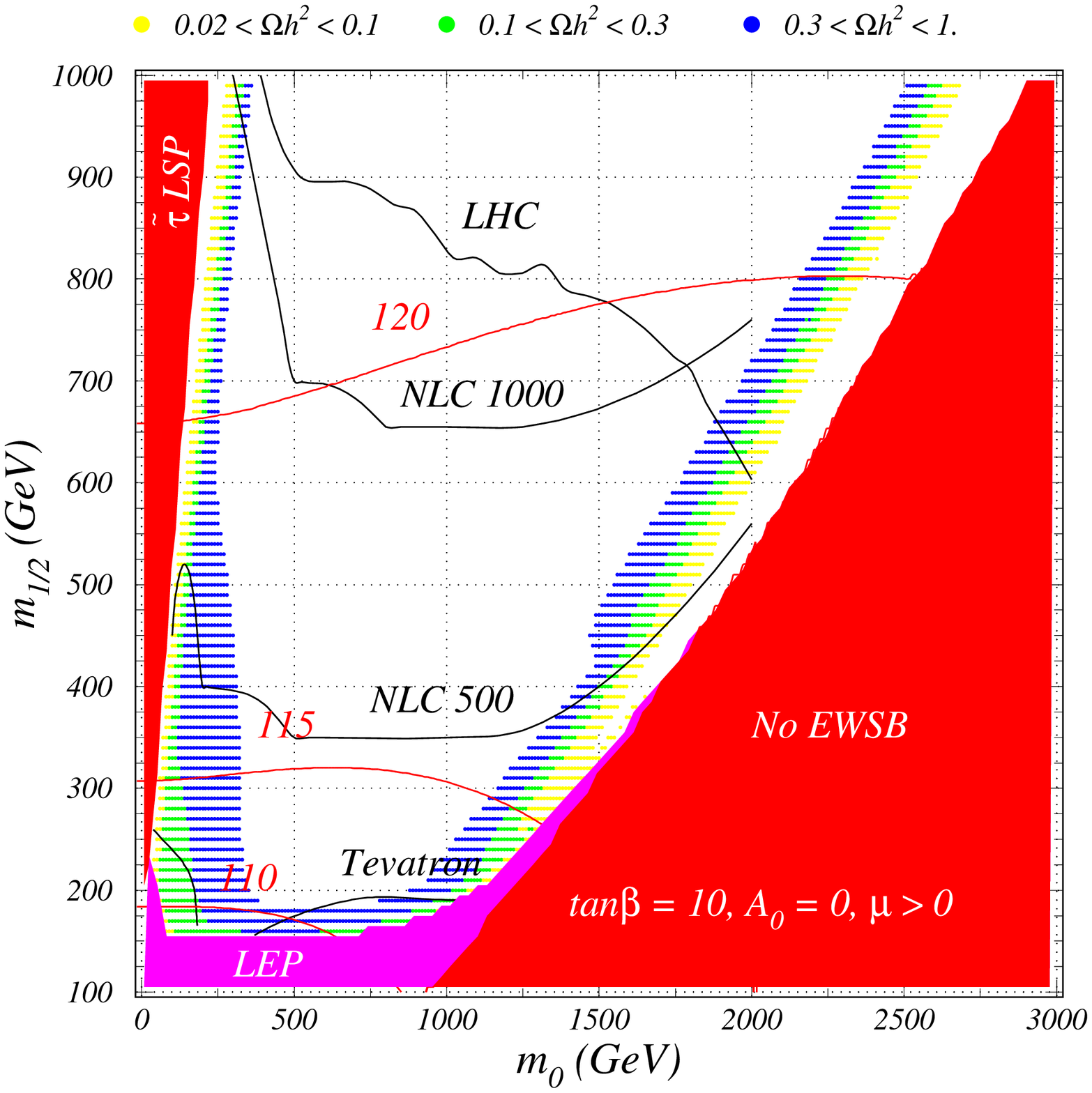,width=15cm} 
        \caption{Regions of relic density in the
        $m_0\ vs.\ m_{1/2}$ plane, including theoretical and
        experimental constraints, contours of light Higgs mass $m_h$ (red),
        and reach projections for the Fermilab Tevatron, CERN LHC
        and Next Linear Collider. We adopt $\tan\beta =10$, $A_0=0$
        and $\mu >0$.}
	\label{concl_10}}

A similar comparison of neutralino relic density versus collider searches
is shown in Fig. \ref{concl_45}, although in this case, results for
the NLC reach are unavailable. A large part of the green region is
actually excluded by the LEP2 Higgs search results. Furthermore, 
the reach of the Fermilab Tevatron barely extends to the cosmologically 
favored
region. The CERN LHC covers most of the green region, with the exception
of the stau co-annihilation band, and the higgsino-like LSP band.

\FIGURE[t]{\epsfig{file=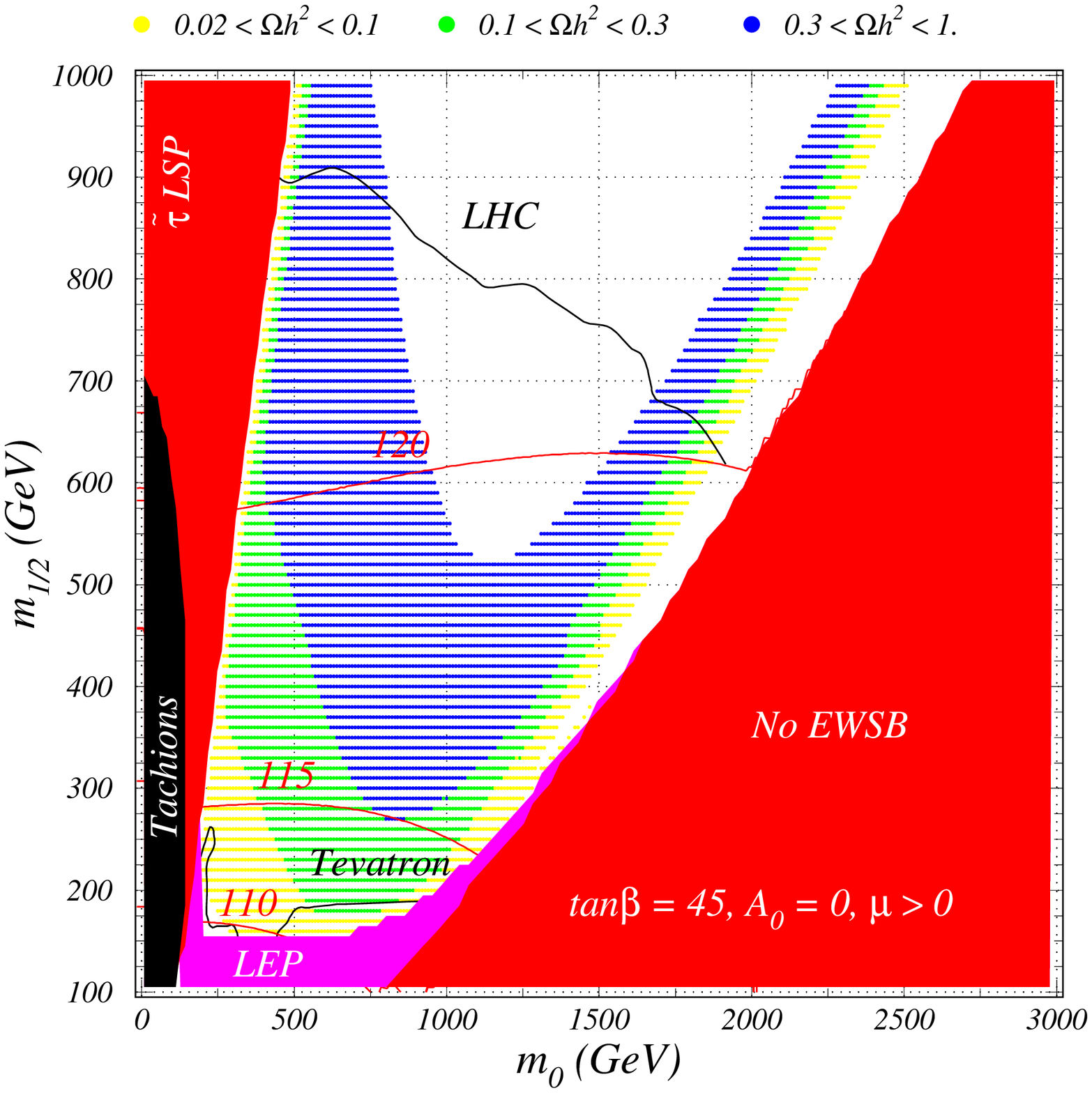,width=15cm} 
        \caption{Regions of relic density in the
        $m_0\ vs.\ m_{1/2}$ plane, including theoretical and
        experimental constraints, contours of light Higgs mass $m_h$ (red),
        and reach projections for the Fermilab Tevatron and CERN LHC.
        We adopt $\tan\beta =45$, $A_0=0$ and $\mu >0$.}
	\label{concl_45}}

\section{Conclusions}

In conclusion, we have performed a calculation of the neutralino
relic density in the minimal supergravity model including all
$2\to 2$ neutralino annihilation and co-annihilation processes, where
the initial state includes $\tz_1$, $\tz_2$, $\tw_1$, $\te_1$, $\tmu_1$,
$\ttau_1$, $\tst_1$ and $\tb_1$. The calculation was performed using the
CompHEP program for automatic evaluation of Feynman diagrams, coupled
with ISAJET for sparticle mass evaluation in the mSUGRA model, and
for standard and supersymmetric couplings and decay widths. 
We implemented relativistic thermal averaging,
which is especially important for evaluating the relic density
when resonances in the annihilation cross section are present, and
neutralino thermal velocities can be relativistic. The three-dimensional
integration was
performed by Monte Carlo evaluation with importance sampling, which
yields in general good convergence even in the presence of
narrow resonances. We note once again that a calculation of
similar scope and procedure was recently reported in Ref. \cite{belanger}.

It may be useful to compare our results against other recent 
evaluations of the neutralino relic density including all 
co-annihilation effects. We compared against several recently
published results\cite{ellis_co,ellis,leszek2,manuel}. Our results agree
qualitatively with these other published results. Quantitatively, these
various papers largely disagree amongst themselves and with our work at 
very large values of $\tan\beta$. In this region, even one loop
calculations of sparticle and Higgs boson masses can be very unstable,
and especially the value of $m_A$ is very sensitive to the exact
procedure involved in calculating sparticle masses. This results in
differences in the precise location of the corridor of annihilation through 
the $A$ and $H$ resonances. Our value of $m_A$ at large $\tan\beta$
seems generally larger than the values obtained in 
Refs. \cite{ellis_co,ellis}, and somewhat smaller than those obtained in
Refs. \cite{leszek2} and \cite{manuel}. Clearly, more theoretical work
needs to be done to stabilize the SUSY and Higgs particle mass 
predictions at large $\tan\beta$.
Finally, the width of the bands of cosmologically favored relic density
around the $A$, $H$ annihilation corridors appears much wider in our
results than in
the results of Refs. \cite{ellis,ellis_co}. This might be an effect of our
improved treatment including {\it relativistic } thermal averaging. It would
be useful to have a comparison against similar results from the
group Belanger et al.\cite{belanger} when these become available.

We presented all our results within the framework of the mSUGRA model.
We found four regions of parameter space that led to relic densities
in accord with results from cosmological measurements, {\it i.e.}
$0.1<\Omega_{\tz_1}h^2<0.3$. These include {\it i}.) the region
dominated by $t$-channel slepton exchange, {\it ii.}) the region
dominated by stau co-annihilation, {\it iii.}) the large $m_0$ region
dominated by a more higgsino-like neutralino and {\it iv.}) the
broad regions at high $\tan\beta$ dominated by off-shell 
annihilation through the $A$ and $H$ Higgs boson resonances.
Regions {\it ii.}) and {\it iii.}) generally have large
fine-tuning associated with them, and although it is logically
possible that nature has chosen such parameters, any slight deviation
of model parameters would lead to either too low or too high a relic
density. Region {\it i.}) generally has the property that some of the
sleptons have masses less than about 300-400 GeV. This region can give 
rise to a rich set of collider signatures, since many of the sparticles
are relatively light. 

Region {\it iv.}) gives broad regions of 
model parameter space with reasonable values of relic density as well
as low values of the fine-tuning parameter. 
It can also allow quite heavy values of SUSY particle masses, 
which would be useful to suppress many flavor-violating 
(such as $b\to s\gamma$)\cite{bsg} and CP violating
loop processes, and the muon $g-2$ value\cite{gm2}. 
In many respects region {\it iv.}) is a favored region of parameter
space. The neutralino relic density may well point the way to the sort of
SUSY signatures we should expect at high energy collider experiments.

\bigskip

\acknowledgments

We thank Manuel Drees, Konstantin Matchev, Leszek Roszkowski 
and Xerxes Tata for discussions.
This research was supported in part by the U.S. Department of Energy
under contract number DE-FG02-97ER41022.

\appendix


\end{document}